\documentclass[twocolumn]{elsart}

\usepackage{natbib}

 \usepackage{graphics}
 \usepackage{graphicx}
 \usepackage{epsfig}

\usepackage{amssymb}

\def\La1215 {Ly$\alpha\lambda1215$}

\begin{document}

\begin{frontmatter}



\title{Magnetic Field Geometry of the Broad Line Radio Galaxy 3C111} 


\author[label1]{P. Kharb},
\ead{rhea@iiap.ernet.in}
\author[label2]{D. Gabuzda},
\author[label4]{W. Alef},
\author[label4]{E. Preuss},
\author[label1]{P. Shastri}


\address[label1]{Indian Institute of Astrophysics, Koramangala P.O., 
Bangalore - 560034}
\address[label2]{Physics Department, University College Cork, Cork, Ireland
\emph{and} Astro Space Center, Lebedev Physical Institute, Moscow, Russia}
\address[label4]{Max-Planck-Institut-fur-Radioastronomie, Bonn, Germany}


\begin{abstract}
Very Long Baseline Polarimetric observations of the Broad Line 
Radio galaxy 3C111 performed in July and September of 1996 at 8 and 43 GHz 
reveal rapidly evolving parsec-scale radio structure after a large
millimetre outburst. The {\bf B}-field geometry is not simple.  We 
present a first analysis of possible Faraday and optical depth effects
based on a comparison of the polarization images for the two frequencies. 
\end{abstract}

\begin{keyword}
Galaxies: Individual (3C111) \sep Interferometry \sep Polarimetry

\end{keyword}

\end{frontmatter}

\section{Introduction}
\vspace*{-0.5cm}
The broad-line radio galaxy 3C111 (0415+739) is
the nearest ($z=0.0485$) 
classical FR~II radio galaxy with a strong compact core at cm/mm 
wavelengths \citep{Wills75}.
Radio observations on kiloparsec scales show a highly collimated jet
leading from the core to the northeastern lobe \citep{LinfieldPerley84}.
On parsec scales, the jet is much more prominent and is one-sided
towards the northeast, roughly aligned with the kpc-scale jet 
\citep{Linfield81}. 

Following a large mm-outburst, we observed the source with the NRAO VLBA 
(10 $\times$ 25~m) and the Effelsberg antenna (100~m) in dual-polarization 
mode at 8.4~GHz and 43~GHz, on July 8 and September 19, 
1996.  
A preliminary analysis of the total intensity images was presented 
by \citet{Alef98}.
We present here the first results of our polarization analysis. 

\vspace*{-0.5cm}
\section{Observations and Results}
\vspace*{-0.5cm}
The data for both epochs were reduced using standard techniques in the
NRAO AIPS package. The instrumental polarizations (`D-terms') were 
determined using the task LPCAL, using the compact source 0420--014 at
43~GHz and the unpolarized source 3C84 at 8~GHz. 
We calibrated the absolute values of the polarization position
angles $\chi$ 
by applying the calibrations determined 
for other VLBA experiments at 43 and 8 GHz within a few 
months of our epochs, taking care to use the same reference
antenna during the calibration (Los Alamos). 
This procedure is justified by the fact that the right--left
phase differences of the VLBA antennas are stable on time scales of
six months or more \citep[e.g.][]{Reynolds01}. 

We show the two 43-GHz and the two 8-GHz images convolved with the same 
beams, corresponding roughly to the beam sizes obtained for uniform 
weighting: $0.20 \times 0.15$~mas in $PA = -20^{\circ}$ for the 43-GHz 
maps and $1.15 \times 0.70$~mas in $PA = -20^{\circ}$ for the 8-GHz maps.
The maps show contours of total intensity ($I$) increasing in steps of 
two with polarization ($P$) vectors superimposed.

\vspace*{-0.5cm}
\subsection{July 8, 1996} 
\vspace*{-0.5cm}
Figure~1a shows the 43~GHz image for July 8, 1996.
The jet to the northeast is clearly visible.
Polarization is reliably detected from the core and a bright knot 
at a distance of $r\sim$ 0.5~mas from the core. The 
$P$ vectors in the core are aligned with the inner
jet, but the relationship between $\chi$ for the bright
knot and the local jet direction is unclear.  The degree of polarization 
in the core region is approximately 1\%, and rises to 3--4\% in the bright knot.

\begin{figure}[h]
(a)
\includegraphics[width=6.5cm,height=6.5cm,angle=-90]{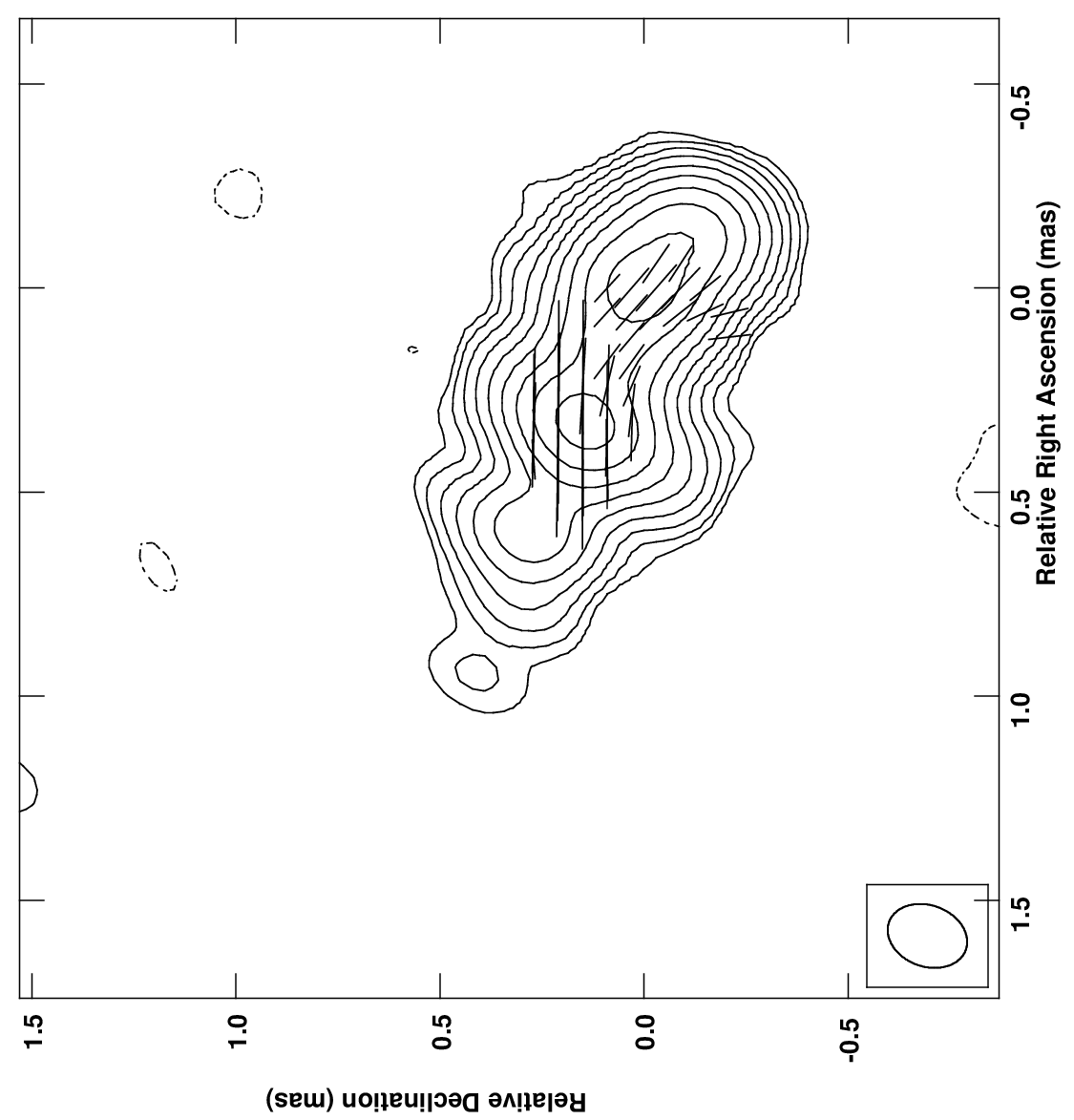}
(b)
\includegraphics[width=6.5cm,height=6.5cm,angle=-90]{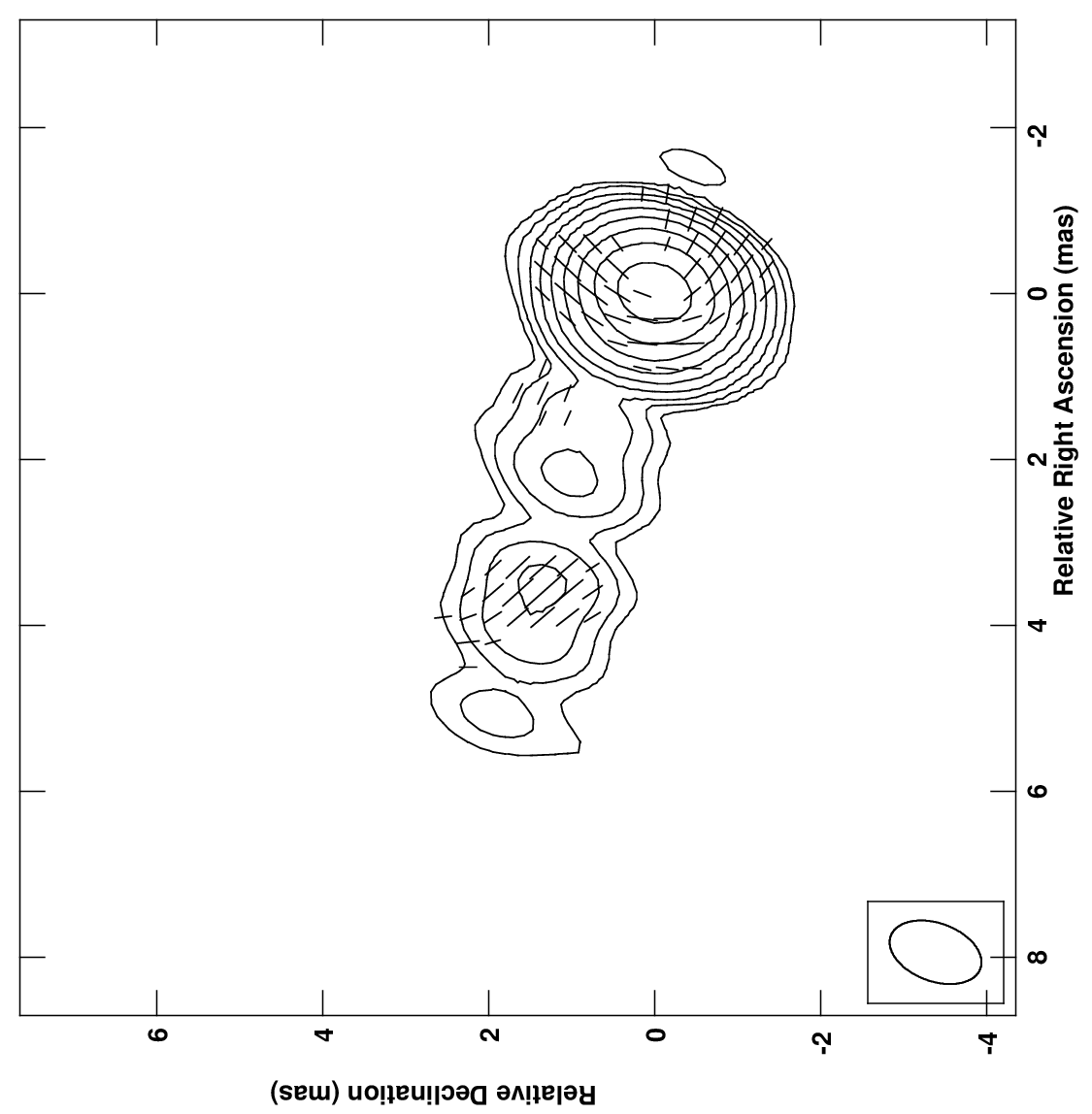}
\caption{$I$ maps of 3C111 for July 8, 1996 at (a) 43 GHz (peak = 
2.12 Jy\, beam$^{-1}$) and (b) 8 GHz (peak = 2.16~Jy\, beam$^{-1}$)
with $P$ vectors superimposed. The bottom contours are $\pm0.25\%$ 
of the peaks.}
\end{figure}

Figure~1b shows the 8~GHz image for the same epoch. The
polarization at the western edge of the core region has roughly
the same $\chi$ as the 43-GHz core 
($\chi \simeq45-55^{\circ}$), suggesting that they may originate
in the same region. 
The inner part of the 8-GHz jet has a region with transverse $P$; 
if Faraday effects are not significant and this region is 
optically thin, this implies a longitudinal {\bf B} field.  However, 
this region roughly coincides with the bright knot at $r\simeq 0.5$~mas
in the 43-GHz image, and we would expect the 43-GHz and 8-GHz
$\chi$'s to be more similar
in the absence of significant Faraday or optical depth effects.
We are in the process of a more detailed analysis 
of the origin of this large offset 
between the observed $P$ vectors at 8~GHz ($\chi\simeq -20^{\circ}$) 
and 43~GHz ($\chi\simeq 90^{\circ}$). 
If Faraday or optical depth effects
are substantial, the 43-GHz $\chi$ more accurately reflects the
underlying {\bf B}-field geometry.  Further from the core 
($r\simeq 4$~mas), the polarization appears to become somewhat 
oblique to the jet direction ($\chi\simeq 45^{\circ}$).

\vspace*{-0.5cm}
\subsection{September 19, 1996}
\vspace*{-0.5cm}
Figure~2a shows the 43~GHz image for September
19, 1996.  
The peak has decreased by a factor of about 1.7. It is now clear 
that the jet first emerges to the northeast, turns 
nearly directly east, then turns again to the northeast. In this light, 
the earlier orientation of the 43-GHz $\chi$ for the knot at
$r \simeq 0.5$~mas now makes sense: $\chi$ was apparently
well aligned with the direction of the jet flow, nearly directly 
eastward. This bright feature has moved to $r \sim 0.75$~mas.  Its 
$\chi$ again appears oblique to the jet direction; however, it is possible 
that it bears some relationship to the local flow direction (either 
parallel or perpendicular to it), but that the flow direction is again not 
known.  The core polarization was not detected at this epoch.

\begin{figure}[h]
(a)
\includegraphics[width=6.5cm,height=6.5cm,angle=-90]{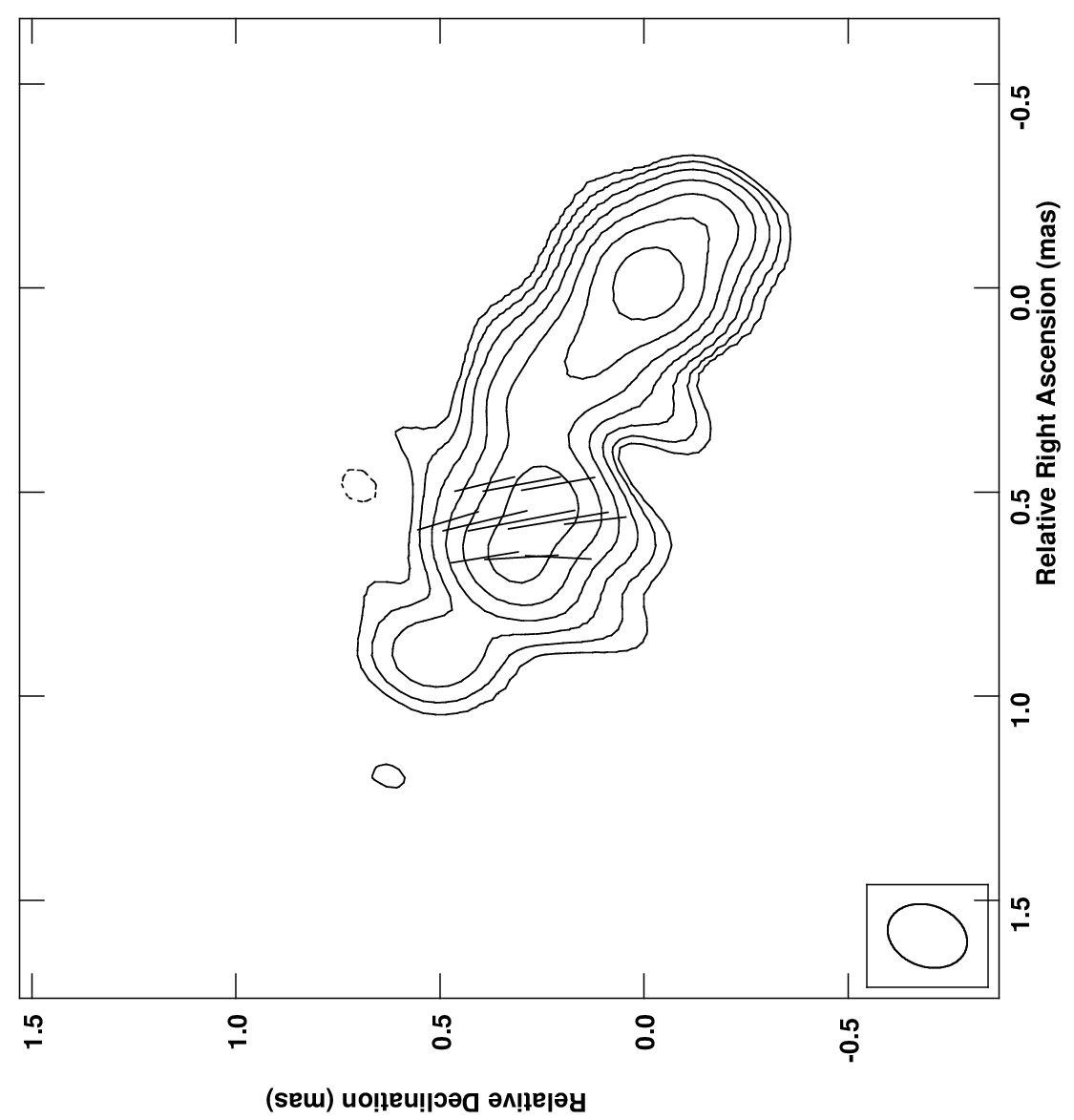}
(b)
\includegraphics[width=6.5cm,height=6.5cm,angle=-90]{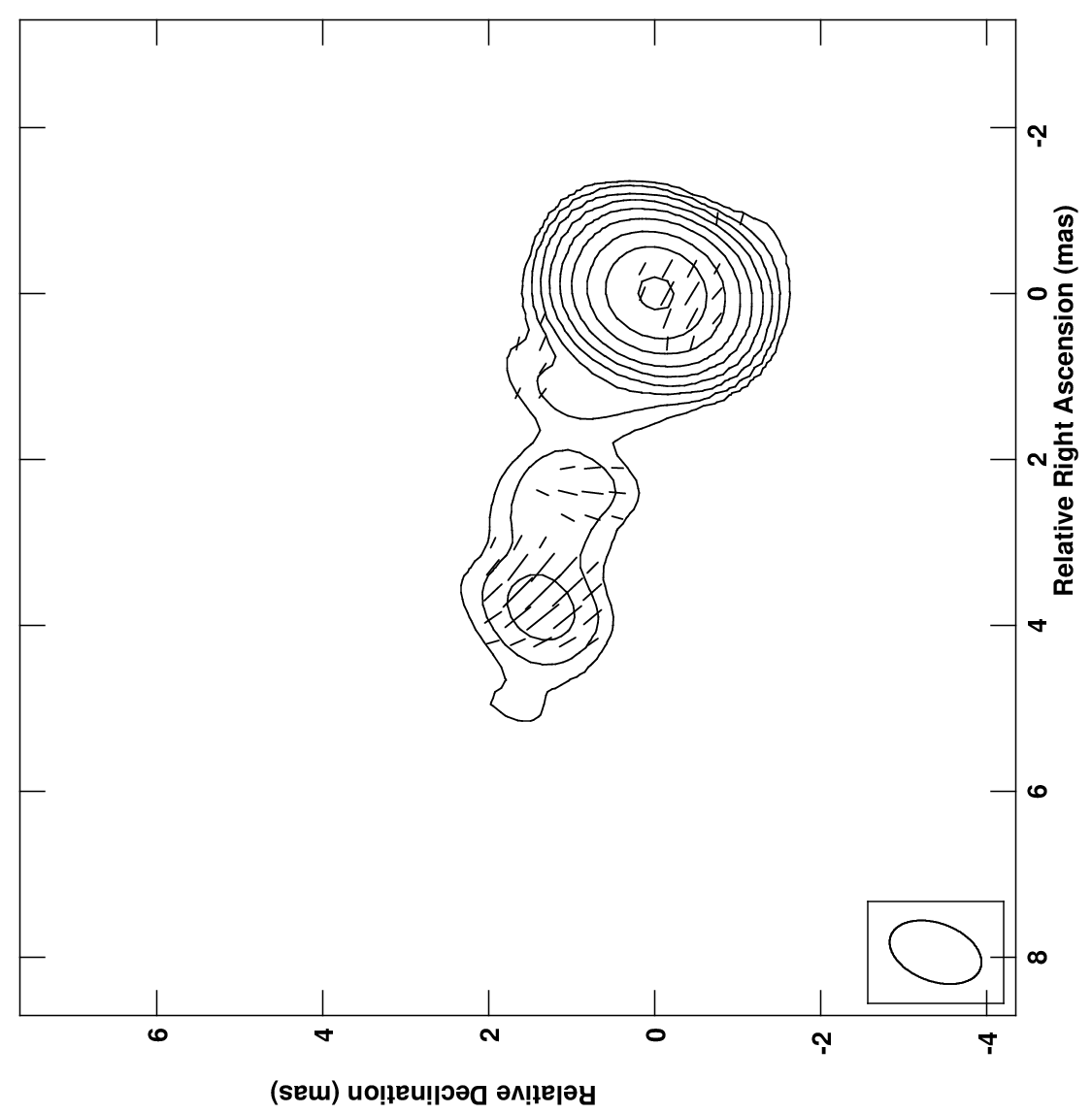}
\caption{$I$ maps of 3C111 for September 19, 1996 at 
(a) 43 GHz and (b) 8 GHz with $P$ vectors superimposed. The bottom 
contours for (a) are $\pm 1.0\%$ of the peak (1.26 Jy\, beam$^{-1}$), and for
(b) are $\pm 0.35\%$ of the peak (2.36~Jy\, beam$^{-1}$).}
\end{figure}

Figure~2b shows the 8~GHz image for the same epoch. 
The polarization in the core region is aligned with the 
inner jet (see Fig.~2a).  There is a region of 
transverse $\chi$ at $r\simeq 3$~mas from the core, implying a longitudinal 
{\bf B} field (if this region is optically thin and Faraday effects are
not important).  The $\chi$ in the knot $r\sim$ 4.5~mas from the core 
is very similar to that at the earlier epoch 
($\chi\simeq 45^{\circ}$), although in both cases this appears to be
somewhat oblique to the local jet direction. The degree of
polarization in the core region is 1--2\%, and rises to 8--10\% in the
3-mas and 4.5-mas knots.
It is interesting that, at both epochs, there is weak polarization at 
the northern edge of the inner 8-GHz jet
with $\chi$ roughly along the jet direction, suggesting this
polarization may be associated with a transverse {\bf B} field that is
not well aligned with the jet ridgeline.

\vspace*{-0.5cm}
\section{Conclusions}
\vspace*{-0.5cm}
Our images of 3C111 for two epochs separated by
only 2.5~months reveal dramatic changes in the VLBI structure after 
a millimetre outburst.  The 43-GHz images show obvious expansion of a 
bright knot from the core. The $\chi$ of this knot rotated 
by about $60-70^{\circ}$ between the two epochs.  At the earlier epoch, 
$\chi$ is well aligned with the local flow, 
implying a transverse {\bf B} field.  The orientation of $\chi$ relative
to the jet flow at the later epoch is not obvious, but may
become clearer as we are able to better elucidate the local flow 
direction.  
The polarization of an 8-GHz knot roughly 
$r\simeq 4$~mas from the core has the same $\chi$ at both epochs, 
somewhat oblique to the jet direction. 

In general, the degree of polarization in the core region is modest 
(1--2\%), while it is somewhat higher in the jet and may increase with 
distance from the core. The degree of polarization reaches 3--4\% at 43-GHz  
$\sim$ 0.5--1~mas from the core and 8--10\% at 8 GHz $\sim$ 3--5~mas 
from the core.

Comparison of our 43 and 8-GHz images for July 1996 shows
a large offset between the $P$ vectors at $r\simeq 0.5-1$~mas, 
indicating that Faraday and/or 
optical-depth effects may be playing a significant role.  However, 
these images also suggest that the $P$ vectors in the 
core region have nearly the same orientations at the two frequencies,
in which case Faraday and optical depths are less important there than
in the inner jet.

We are in the process of a more detailed analysis of these data,
including model fitting of the $I$ and $P$ data.
In addition, we are reducing additional 8+43~GHz VLBA data
for June 1997, as well as 43-GHz data for December 1997,
May 1998, and December 1998. 

\vspace*{-0.5cm}
\section{Acknowledgements}
\vspace*{-0.5cm}
P. Kharb is grateful for local support from the conference organisers and 
travel support from the Council for Scientific \& Industrial Research, 
Govt. of India.  The National Radio Astronomy Observatory is a facility 
of the National Science Foundation operated under cooperative agreement by
Associated Universities, Inc.

\vspace*{-0.5cm}
\bibliographystyle{aa}
\bibliography{p_pkharb}

\end{document}